\documentclass[preprint,showpacs,amsmath,amssymb]{revtex4}
\usepackage{graphicx}
\usepackage{dcolumn}
\usepackage{bm}
\usepackage{psfig}

\newcommand{\pnbpi}{$J/\psi\to p\pi^-\bar n$ }

\begin{document}
\preprint{Draft-PRL}

\title{\boldmath Observation of Two New $N^*$ Peaks in $J/\psi \to p
\pi^- \bar n$ and $\bar p\pi^+n$ Decays}

\author{M.~Ablikim$^{1}$, J.~Z.~Bai$^{1}$, Y.~Ban$^{10}$,
J.~G.~Bian$^{1}$, X.~Cai$^{1}$, J.~F.~Chang$^{1}$,
H.~F.~Chen$^{16}$, H.~S.~Chen$^{1}$, H.~X.~Chen$^{1}$,
J.~C.~Chen$^{1}$
% ...
\\(BES Collaboration)\\
$^1$ Institute of High Energy Physics, Beijing 100039, People's Republic of China\\
% ...
$^{10}$ Peking University, Beijing 100871, People's Republic of China\\
% ...
$^{16}$ University of Science and Technology of China, Hefei 230026, People's Republic of China\\
% ...
}

\date{July 15, 2006}

\begin{abstract}
  The decay $J/\psi\to\bar NN\pi$ provides an effective isospin 1/2
  filter for the $\pi N$ system due to isospin conservation. Using 58
  million $J/\psi$ decays collected with the Beijing Electromagnetic
  Spectrometer (BES) at the Beijing Electron Positron Collider (BEPC),
  more than 100 thousand $J/\psi \to p \pi^- \bar n + c.c.$ events are
  obtained. Besides the two well known $N^*$ peaks at around 1500
  MeV/$c^2$ and 1670 MeV/$c^2$, there are two new, clear $N^*$ peaks
  in the $p\pi$ invariant mass spectrum around 1360 MeV/$c^2$ and 2030
  MeV/$c^2$ with statistical significance of $11\sigma$ and $13\sigma$, respectively.
  We identify these as the first direct observation of the
  $N^*(1440)$ peak and a long-sought ``missing" $N^*$ peak above 2
  GeV/$c^2$ in the $\pi N$ invariant mass spectrum.
\end{abstract}


\maketitle

The nucleon is the simplest system in which the three colors of
QCD can combine to form a colorless object, and the essential
nonabelian character of QCD is manifest. It is necessary to
understand the internal quark-gluon structure of the nucleon and
its excited $N^*$ states
 before we can claim to really understand the strong
interaction.

A very important source of information for the nucleon internal
structure is the $N^*$ mass spectrum as well as its various
production and decay rates. Our present knowledge of these comes
almost entirely from $\pi N$ experiments performed more than
twenty years ago \cite{Eidelman:2004wy}. Because of its importance
for the understanding of nonperturbative QCD, a series of new
experiments on $N^*$ physics with electromagnetic probes (real
photons and electrons with space-like virtual photons) have
recently been started at JLAB, ELSA at Bonn, GRAAL at Grenoble and
SPRING8 at JASRI \cite{Nstar02}. They have already produced some
important results
\cite{Ripani:2002ss,Tran:1998qw,Assafiri:2003mv,Nakano:2003qx}.
Nevertheless, our knowledge of the $N^*$ resonances remains very
poor. Even for the well-established lowest excited state, the
$N^*(1440)$, properties such as mass, width and decay branching
ratios etc., still have large experimental uncertainties
\cite{Eidelman:2004wy}. Another outstanding problem is that, in
many of its forms, the quark model predicts a substantial number
of ``missing" $N^*$ states around 2 GeV/$c^2$, which have not so
far been observed \cite{Capstick:2000qj}. The difficulty in
extracting information on these high mass $N^*$ resonances from
$\pi N$ and $\gamma N$ experiments is the overlap of many broad
resonances with various spin and isospin. Recently $J/\psi$ decays
at the BEPC were proposed \cite{Zou:2000re} as an excellent place
to study $N^*$ resonances, and $N^*$ production from $J/\psi\to
p\bar p\eta$ has been studied based on 7.8 million BESI $J/\psi$
events \cite{Bai:2001ua}.

In this letter, we report on a study of $N^*$ resonances from $J/\psi
\to p \pi^- \bar n$ and $\bar p\pi^+ n$ channels based on 58 million
$J/\psi$ events collected with the BESII detector at the BEPC.  Due to
isospin conservation, the $\pi N$ system in $J/\psi$ decay is pure
isospin 1/2.  Compared with $\pi N$ and $\gamma N$ experiments, which
mix isospin 1/2 and 3/2, this is a big advantage.

BESII is a large solid-angle magnetic spectrometer that is described
in detail in Ref. \cite{Bai:1994zm}. Charged particle momenta are
determined with a resolution of $\sigma_p/p=1.78\%\sqrt{1+p^2}$ (with
$p$ in GeV/c) in a 40-layer cylindrical drift chamber. Particle
identification is accomplished using specific ionization ($dE/dx$)
measurements in the drift chamber and time-of-flight (TOF)
measurements in a barrel-like array of 48 scintillation counters. The
$dE/dx$ resolution is $\sigma_{dE/dx}=8.0\%$, and the TOF resolution
is $\sigma_{TOF}=180$ ps for the Bhabha events.  The combination of
$dE/dx$ and TOF information provides efficient identification of the
charged pion and proton (or antiproton) for the $J/\psi \to p \pi^-
\bar n$ and $\bar p\pi^+ n$ processes.

A GEANT3 based Monte Carlo simulation package
(SIMBES)\cite{Ablikim:2005py} with detailed consideration of real
detector performance (such as dead electronic channels) is used for
estimating the detection efficiency and background contributions. The
consistency between data and Monte Carlo has been carefully checked in
many high purity physics channels, and the agreement is quite
reasonable.

For the decay $J/\psi \to p \pi^- \bar n$, the anti-neutron is not
detected directly. We select the $p$ and $\pi^-$ from two prong events
with oppositely charged tracks and require the missing mass to be
consistent with the $\bar n$ mass.

\begin{figure}[htbp]
  \centering
  \includegraphics[width=0.45\textwidth,height=0.25\textheight]{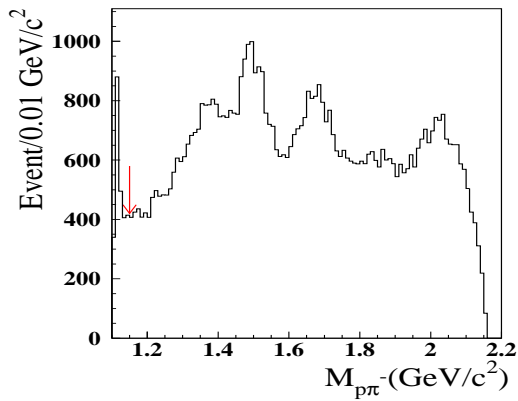}
  \caption{The $p\pi$ invariant mass spectrum before the $M_{p\pi} >
    1.15 {\rm GeV}/c^2$ requirement.}
  \label{ppicut}
\end{figure}

\begin{figure}[htbp]
  \centering
  \includegraphics[width=0.45\textwidth,height=0.25\textheight]{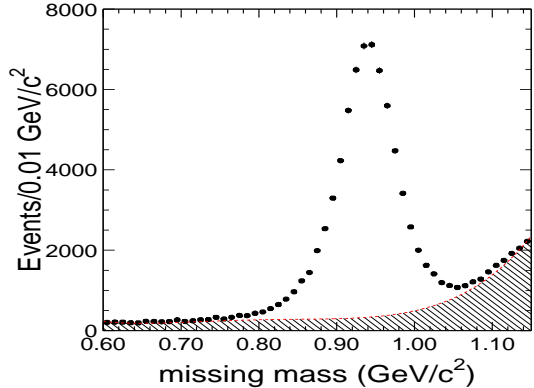}
  \caption{The missing mass distribution for $J/\psi\to p\pi^-\bar n$
    with the fitted background shown by the shaded area.}
  \label{miss}
\end{figure}

In the event selection, we require each charged track to be well
fitted to a helix originating near the interaction point and to be
within the polar angle region $|\cos\theta|< 0.8$. Both the $dE/dx$
and TOF information are used to form particle identification
confidence levels $\wp^i_{pid}$, where $i$ denotes $\pi$, $K$, or $p$.
For the positive charged track, we require $\wp^p_{pid}>\wp^K_{pid}$
and $\wp^p_{pid}>\wp^\pi_{pid}$; for the negative charged track, we
require $\wp^\pi_{pid}>\wp^K_{pid}$ and $\wp^\pi_{pid}>\wp^p_{pid}$.
After these requirements, the $p\pi^-$ invariant mass spectrum shown
in Fig.~\ref{ppicut} is obtained. There is a narrow $\Lambda$ peak at
1.116 GeV due to background channels with $\Lambda$s. After imposing a
further requirement of $M_{p\pi} > 1.15 {\rm GeV}/c^2$ to remove this
background, the missing mass distribution shown in Fig.~\ref{miss} is
obtained.  A clear $\bar n$ peak is observed.

A similar selection is used for the charge conjugate channel
$J/\psi\to\bar p\pi^+n$. The missing mass distribution of $\bar
p\pi^+$ is almost identical with that for $J/\psi\to p\pi^-\bar n$.
Fitting the missing mass spectra between 0.60 and 1.15 GeV/$c^2$ with
signal plus background described by a fourth order Chebyshev
polynomial and using Monte-Carlo determined acceptances for these two
channels, branching ratios for these two channels are measured as
$(2.36\pm 0.02\pm 0.21)\times 10^{-3}$ and $(2.47\pm 0.02\pm
0.24)\times 10^{-3}$ for $p\pi^-\bar n$ and $\bar p\pi^+ n$,
respectively. The statistical errors are small due to the large data
sample. Systematic errors come from several sources. Background
uncertainty is about 4\%. The difference of particle identification
between data and Monte Carlo is about 5\%.  Uncertainty on the total
number of $J/\psi$ events is $\sim 4.7\%$, and the tracking error is
$\sim 4\%$. Combining these errors in quadrature gives a total
systematic error of $(9-10)\%$.

\begin{figure}[htbp]
\centering
\includegraphics[width=8cm]{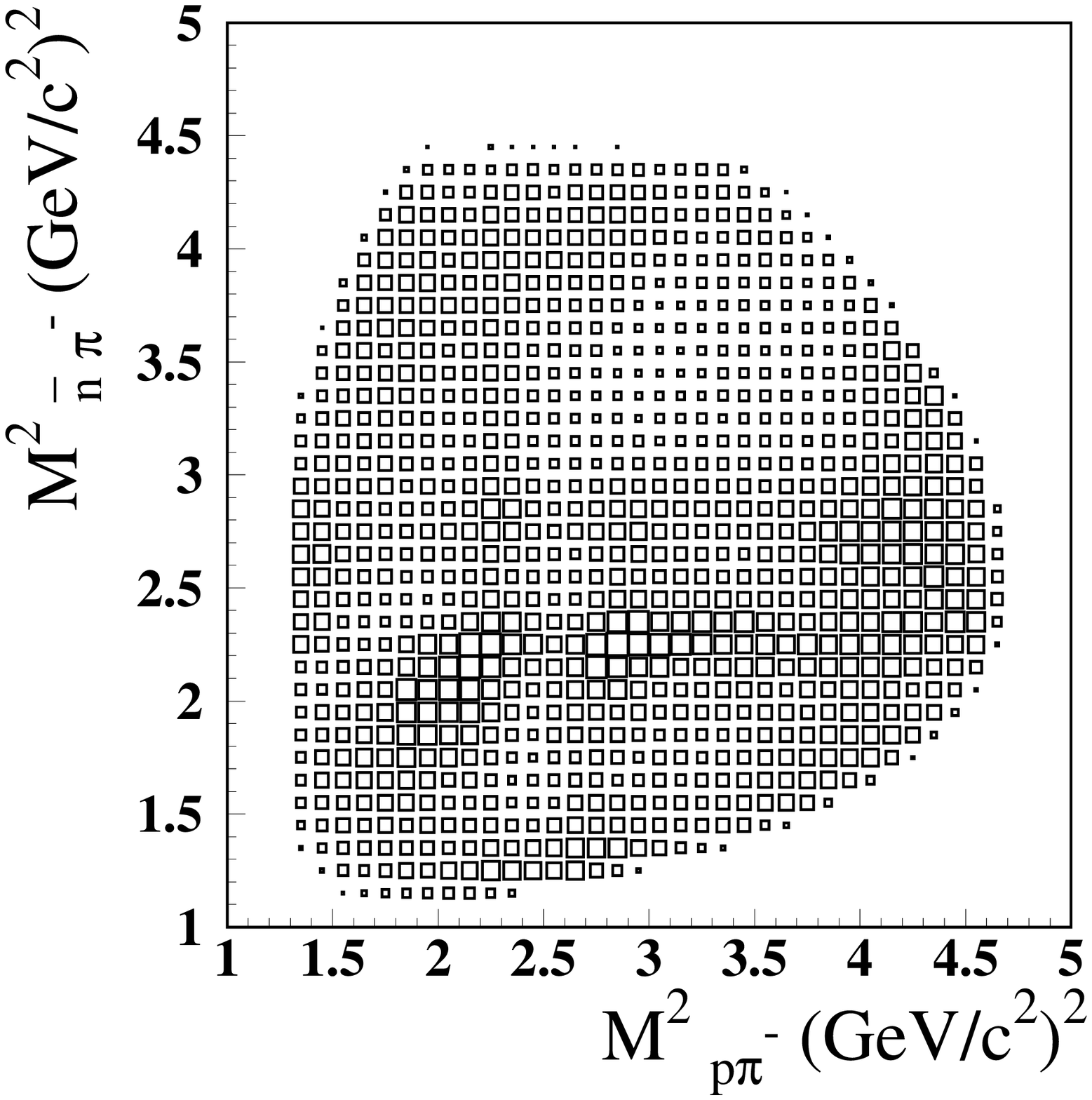}
\includegraphics[width=8cm]{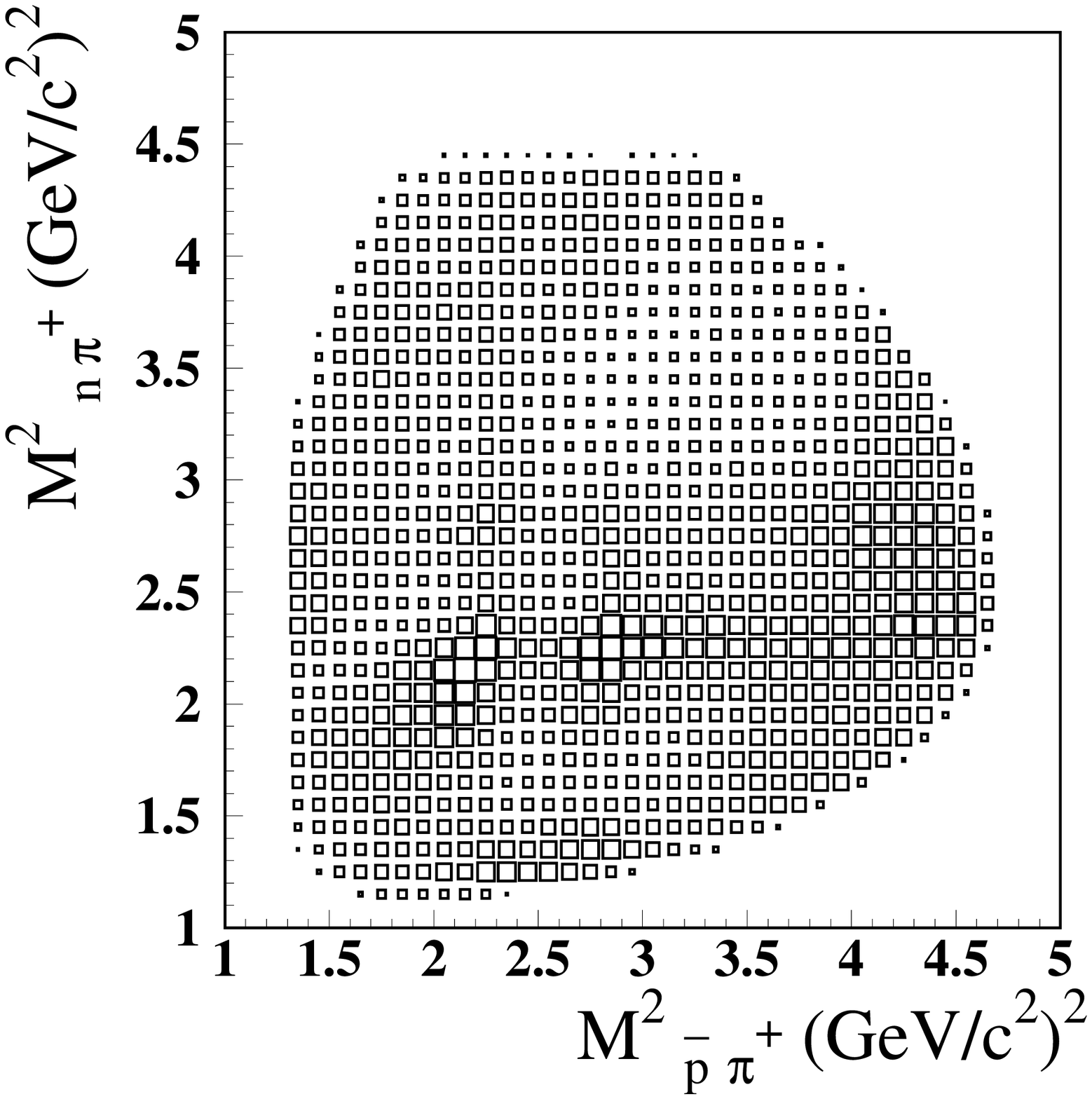}
\caption{Dalitz plots of $M^2_{n\pi}$ vs. $M^2_{p\pi}$ for $J/\psi\to
  p\pi^-\bar n$ (left) and $\bar p\pi^+n$ (right).}
\label{dalitz}
\end{figure}
\begin{figure}[htbp]
    \includegraphics[width=8.5cm,height=7.0cm]{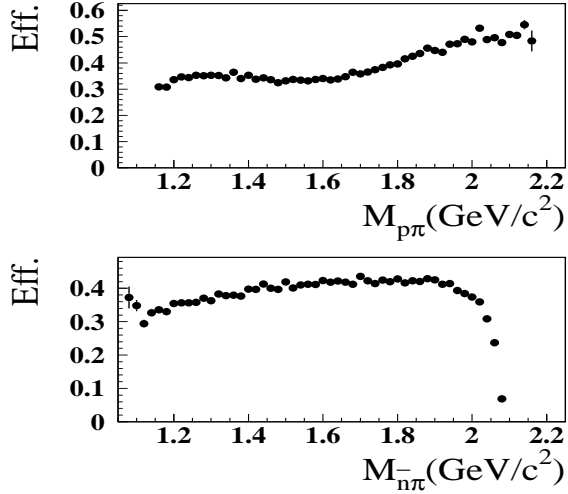}
    \caption{The selection efficiency versus invariant mass for $p\pi$
      (upper plot) and $\bar n\pi$ (lower plot) in the decay \pnbpi.}
    \label{fig:mNpieff}
\end{figure}

Taking $\bar n$/$n$ events in the region of the missing mass
distribution within $0.94\pm 0.06$ GeV/$c^2$, we get 58822 and 54524
events for $J/\psi\to p\pi^-\bar n$ and $\bar p\pi^+ n$, respectively,
including about 6\% background. The Dalitz plots for these two
channels are shown in Fig.~\ref{dalitz}, and are very similar. The
asymmetry between $p\pi$ and $n\pi$ is partly due to differences in
detection efficiency and partly due to isospin symmetry breaking
effects from the electromagnetic interaction.  The $p\pi$ and $n\pi$
selection efficiencies versus invariant mass in the decay $J/\psi\to
p\pi^-\bar n$ are shown in Fig.~\ref{fig:mNpieff}. The efficiency has
a sharp drop for $\bar n\pi^-$ invariant mass above 2 GeV/$c^2$, where
the recoil proton has low momentum and BESII detection efficiency is
correspondingly low.

\begin{figure}[htbp]
\centering
\includegraphics[width=8cm]{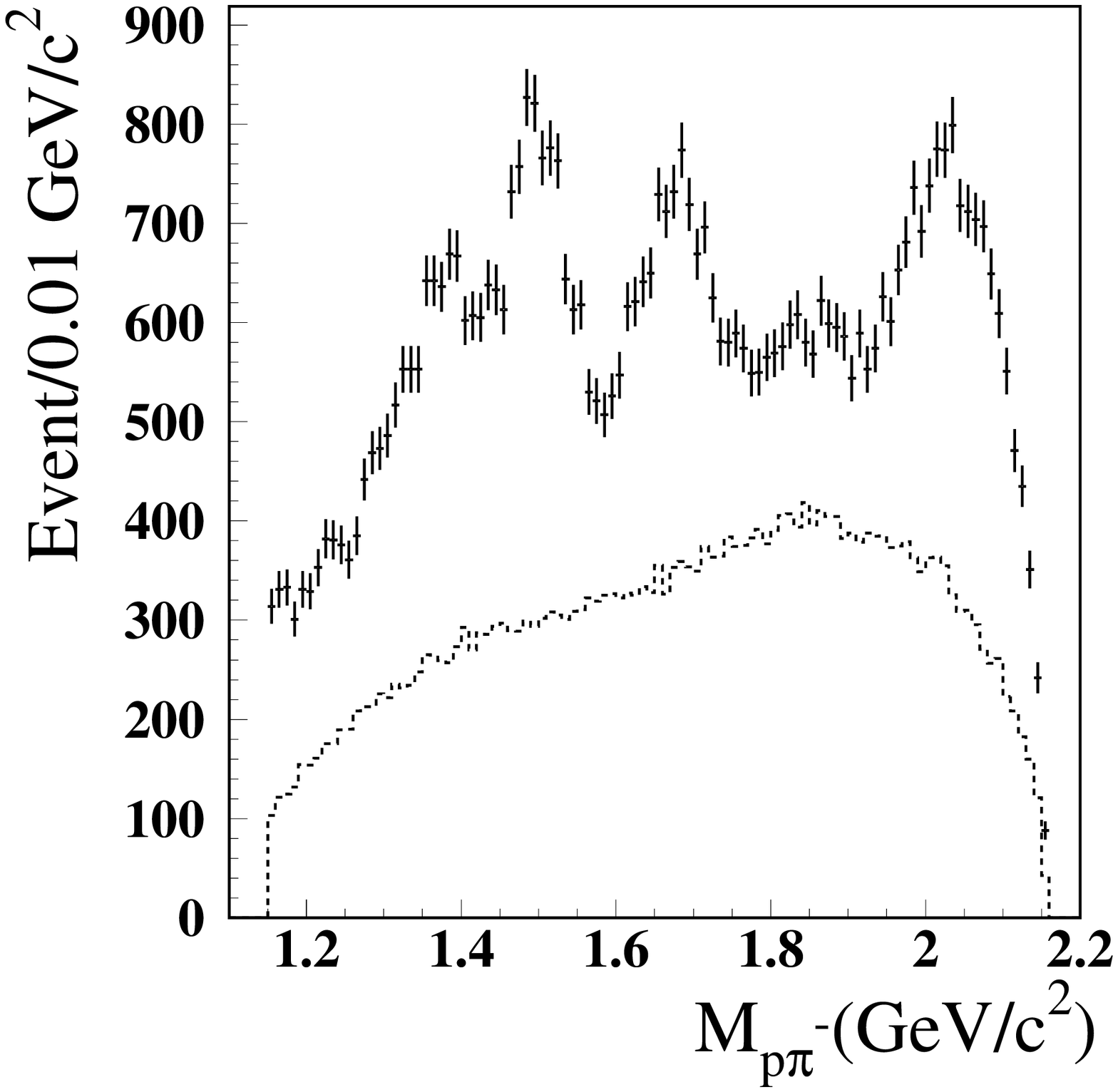}
\includegraphics[width=8cm]{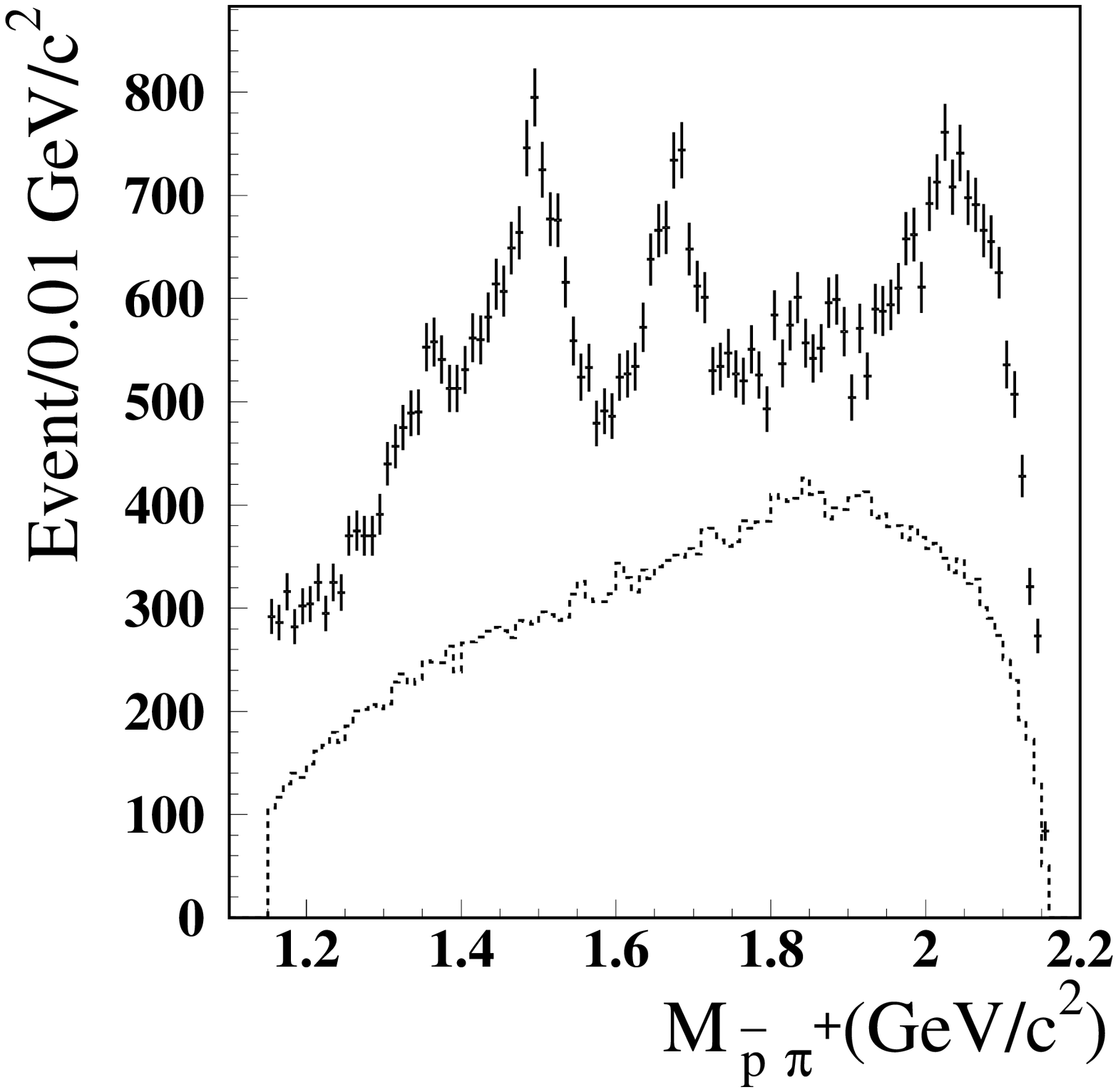}
\caption{ The $p\pi^-$ and $\bar p\pi^+$ invariant mass spectra for
  $J/\psi\to p\pi^-\bar n$ (left) and $\bar p\pi^+ n$ (right),
  compared with phase space.} \label{invm}
\end{figure}

Shown in Fig.~\ref{invm} are the $p\pi^-$ and $\bar p\pi^+$ invariant
mass spectra, as well as the phase space distributions (times
detection efficiency) obtained with the SIMBES Monte Carlo.  The two
invariant mass spectra look very similar.

\begin{figure}[htbp]
  \centering
  \includegraphics[width=9.0cm,height=9.0cm]{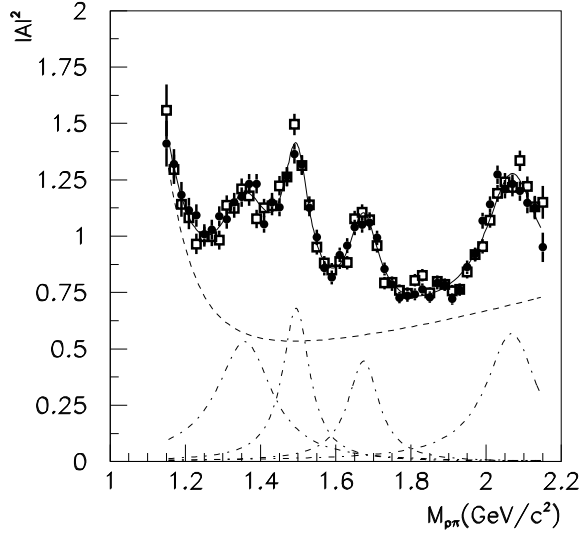}
  \caption{Data divided by Monte Carlo phase space versus $p\pi$
    invariant mass for $J/\psi\to p\pi^-\bar n$ (solid circle) and
    $J/\psi\to\bar p\pi^+n$ (open square), compared with our fit
    (solid curve). The contributions of each resonance peak are shown
    by the dot-dashed lines in the same figure. The dashed line is the
    contribution of background terms including the nucleon pole term.
  }
  \label{fig:vpin}
\end{figure}

To investigate the amplitude squared behavior as a function of
invariant mass, we remove the phase space factor and efficiency
factor from the invariant mass distribution by dividing the data
by Monte Carlo phase space times the detection efficiency. The
results are shown in Fig.~\ref{fig:vpin} with data and Monte Carlo
for the two channels normalized to the same number of events as
for $p\pi^-\bar n$ data. The results for the two charge conjugate
channels are consistent. At low $p\pi$ invariant mass, the tail
from the nucleon pole term, expected from theoretical
considerations \cite{Sinha:1984qn,Liang:2004sd}, is clearly seen.
There are clearly four peaks around 1360, 1500, 1670 and 2065
MeV/$c^2$.  Note that the well known first resonance peak
($\Delta(1232)$) in $\pi N$ and $\gamma N$ scattering data does
not show up here due to the $J/\psi$ decay isospin filter. While
the two peaks around 1500 MeV/$c^2$ and 1670 MeV/$c^2$ correspond
to the well known second resonance peak and the third resonance
peak observed in $\pi N$ and $\gamma N$ scattering data, the two
peaks around 1360 MeV/$c^2$ and 2065 MeV/$c^2$ have never been
observed before in $\pi N$ invariant mass spectra, although a peak
around 1400 MeV was observed in the missing mass spectrum in
$pp\to p + X$ reaction \cite{Morsch:2005cw} with $X$ containing
not only $\pi N$ but also $\pi\pi N$, $3\pi N$, etc.. The one
around 1360 MeV/$c^2$ should be from $N^*(1440)$ which has a pole
around 1360 MeV/$c^2$
\cite{Eidelman:2004wy,Arndt:2003if,Vrana:1999nt} and which is
usually buried by the strong $\Delta$ peak in $\pi N$ and $\gamma
N$ experiments; the other one around 2065 MeV/$c^2$ may be one or
more of the long sought ``missing" $N^*$ resonance(s).

To estimate the mass and width of these resonance peaks, we use simple
relativistic Breit-Wigner functions plus a smooth function to fit the
data points of Fig.~\ref{fig:vpin}. For the final fit we use

\begin{equation}
  \label{eq:rbw}
  |A|^2 = |C_0+C'_0s_{\pi p}|+ \sum_{i=1}^5\frac{|C_i|}{(s_{\pi p}-M_i^2)^2+M_i^2\Gamma_i^2}
\end{equation}
where $s_{\pi p}=M^2_{\pi p}$ is the invariant mass squared of the
$p\pi$ system. Background is described with a constant and $s_{\pi
p}$ linear term plus an additional Breit-Wigner at low energy to
simulate the contribution from the nucleon pole term and the
$\Lambda$ and other backgrounds.  All parameters are free. The
fitted mass and width for the four $N^*$ peaks are listed in Table
\ref{tab:mppifit}. The systematic error in the table is obtained
from the variation of fit results when assuming other background
functions, such as $|C_0|$, $|C_1M_{\pi p}|$, $|C'_0s_{\pi p}|$,
$|C_0+C_1M_{\pi p}|$, $|C_0+C_1M_{\pi p}+C'_0s_{\pi p}|$, and
adding one more $N^*$ resonance around 1.85 GeV/$c^2$ or above 2.2
GeV/$c^2$ in the fitting. The fit to the data is also shown in
Fig. \ref{fig:vpin} with $\chi^2=133$ for 102 data points. All
four peaks are highly significant.  The statistical significance
is  $11\sigma$ for the $N^*(1440)$ peak (the least significant
one) and $13\sigma$ for the new $N^*(2065)$ peak.

\begin{table}[htbp]
  \centering
  \caption{The fitted masses and widths for the four $N^*$ peaks shown in
  Fig.~\ref{fig:vpin}. }
  \begin{tabular}{|c|c|}
    \hline
    Mass(MeV/$c^2$) & Width(MeV/$c^2$)  \\ \hline
    $1358\pm 6\pm 16$ & $179\pm 26 \pm 50$  \\
    $1495\pm 2\pm 3$ & $87\pm 7\pm 10$  \\
    $1674\pm 3\pm 4$ & $100\pm 9\pm 15$  \\
    $2068\pm 3^{+15}_{-40}$ & $165\pm 14\pm 40$ \\ \hline
\end{tabular}
  \label{tab:mppifit}
\end{table}

Because we use a constant width for the Breit-Wigner (BW) formulae,
the BW mass and width are very close to their corresponding pole
positions. For the two well-known peaks at 1500 MeV/$c^2$ and 1670
MeV/$c^2$, the former contains two well-established $N^*$ resonances,
the $N^*(1520)$ and $N^*(1535)$, while the latter contains more than
two $N^*$ resonances. Here we only use one BW function to fit each of
them.

For the new $N^*(2065)$ peak, orbital angular momentum $L=0$ is
preferred due to the suppression of the centrifugal barrier factor for
$L\geq 1$. If we fit the $N^*(2065)$ peak in Fig.~\ref{fig:vpin} with
$L=1$ centrifugal barrier factor instead of Eq.(1), then the $\chi^2$
increases from 133 to 163 for 102 data points. The much worse fit with
$L=1$ compared with $L=0$ means that there is substantial $L=0$
component for the new $N^*(2065)$ peak.

For $L=0$, the spin-parity of $N^*(2065)$ is limited to be $1/2^+$ and
$3/2^+$. This may be the reason that the $N^*(2065)$ shows up as a
peak in $J/\psi$ decays while no peak shows up in the $\pi N$
invariant mass spectra in $\pi N$ and $\gamma N$ production processes
which allow all $1/2^{\pm}$, $3/2^{\pm}$, $5/2^{\pm}$, and $7/2^{\pm}$
$N^*$ resonances and their isospin 3/2 $\Delta^*$ partners around this
energy to interfere with each other. In order to determine its
spin-parity, a partial wave analysis using an effective Lagrangian
approach \cite{Bai:2001ua} is tried for the $p\pi^-\bar n$ data by
including a new $N^*(2065)$ with spin-parity either $1/2^+$ or $3/2^+$
in addition to all well-established $N^*$ resonances below 2 GeV/$c^2$
with masses and widths fixed to their PDG values
\cite{Eidelman:2004wy}. Comparing with the fit without including any
new $N^*(2065)$ resonance, including a $N^*(2065)$ with spin-parity of
either $1/2^+$ or $3/2^+$ improves log likelihood value by more than
400. The new $N^*(2065)$ peak cannot be reproduced by reflections of
well-established $N^*$ resonances. However, the spin-parity of the new
resonance(s) cannot be well determined. The difference of the fits
with different spin-parity quantum numbers is small and depends on
many fitting details, such as how to treat the background contribution
and how large an isospin breaking effect is allowed. Including both
$1/2^+$ and $3/2^+$ improves log likelihood value further by more than
60. So it is quite possible that both are needed. There are quark
model predictions for the existence of $N^*$ resonances with
spin-parity $1/2^+$ and $3/2^+$ in the energy range from 2.0 to 2.1
GeV/$c^2$ \cite{Capstick:2000qj,Isgur:1978wd,Capstick:1992th}. The
same channel has also been studied in $\psi'$ decays
\cite{Ablikim:2006ah}. However due to lower statistics and the
allowance of more partial waves, only a much broader peak was observed
above 2 GeV/$c^2$.

In summary, the $J/\psi\to\bar NN\pi$ decay at BEPC provides an
excellent place for studying pure isospin 1/2 excited $N^*$
resonances. Using 58 million $J/\psi$ decays, more than 100
thousand $J/\psi \to p \pi^- \bar n + c.c.$ events are obtained.
The corresponding branching ratios are determined to be $(2.36\pm
0.02\pm 0.21)\times 10^{-3}$ and $(2.47\pm 0.02\pm 0.24)\times
10^{-3}$ for $p\pi^-\bar n$ and $\bar p\pi^+ n$, respectively. In
the $p\pi$ invariant mass, besides the two well-known peaks around
1500 MeV/$c^2$ and 1670 MeV/$c^2$, there are two new clear peaks.
The one around 1360 MeV/$c^2$ with statistical significance of
$11\sigma$, identified as due to the $N^*(1440)$ resonance
(usually obscured by the strong $\Delta$ peak in $\pi N$ and
$\gamma N$ experiments), has a Breit-Wigner mass and width of
$1358\pm 6 \pm 16$ MeV/$c^2$ and $179\pm 26\pm 50$ MeV/$c^2$. The
other $N^*$ peak around 2065 MeV/$c^2$ with statistical
significance of $13\sigma$ is believed to be due to long sought
``missing" $N^*$ resonance(s) predicted by many theoretical quark
models \cite{Capstick:1992th} in this energy range. A simple
Breit-Wigner fit gives its mass as $2068\pm 3^{+15}_{-40}$
MeV/$c^2$ and width as $165\pm 14\pm 40$ MeV/$c^2$. From a partial
wave analysis, we conclude that the new $N^*(2065)$ peak cannot be
reproduced by reflections of PDG established $N^*$ resonances and
includes at least one new $N^*$ resonance.

\vspace {1cm}

The BES collaboration thanks the staff of BEPC for their hard efforts,
and D.V.Bugg, V.Burkert, H.C.Chiang, S.Dytman, H.T.S.Lee, M.Manley and
P.N.Shen for useful discussions. This work is supported in part by the
National Natural Science Foundation of China under contracts Nos.
19991480, 10225524, 10225525, the Chinese Academy of Sciences under
contract No. KJ 95T-03, the 100 Talents Program of CAS under Contract
Nos. U-11, U-24, U-25, and the Knowledge Innovation Project of CAS
under Contract Nos.  KJCX2-SW-N02, U-602(IHEP); by the National
Natural Science Foundation of China under Contract No. 10175060
(USTC); and by the Department of Energy under Contract No.
DE-FG03-94ER40833 (U Hawaii).

\end{document}